\documentclass[12pt]{article}
\usepackage{graphicx}
\usepackage{amssymb}
\usepackage{amsmath}
\renewcommand{\theequation}{\thesection.\arabic{equation}}

\newlength{\extraspace}
\newlength{\extraspaces}
\newcommand{\be}{\begin{equation}
\addtolength{\abovedisplayskip}{\extraspaces}
\addtolength{\belowdisplayskip}{\extraspaces}
\addtolength{\abovedisplayshortskip}{\extraspace}
\addtolength{\belowdisplayshortskip}{\extraspace}}
\newcommand{\ee}{\end{equation}}

\newcommand{\ba}{\begin{eqnarray}
\addtolength{\abovedisplayskip}{\extraspaces}
\addtolength{\belowdisplayskip}{\extraspaces}
\addtolength{\abovedisplayshortskip}{\extraspace}
\addtolength{\belowdisplayshortskip}{\extraspace}}
\newcommand{\ea}{\end{eqnarray}}

\newcommand{\newsection}[1]{
\vspace{12mm}
\pagebreak[3]
\addtocounter{section}{1}
\setcounter{equation}{0}
\setcounter{subsection}{0}
\noindent{\bf \thesection. #1}
\nopagebreak
\medskip
\nopagebreak}

\newcommand{\newsubsection}[1]{
\vspace{0.8cm}
\pagebreak[3]
\addtocounter{subsection}{1}
\noindent{\it \thesubsection. #1}
\nopagebreak
\vspace{2mm}
\nopagebreak}

\newcounter{saveeqn}
\newcommand{\alpheqn}{\setcounter{saveeqn}{\value{equation}}%
 \stepcounter{saveeqn}\setcounter{equation}{0}%
 \renewcommand{\theequation}
     {\mbox{\thesection.\arabic{saveeqn}\alph{equation}}}}
\newcommand{\reseteqn}{\setcounter{equation}{\value{saveeqn}}%
  \renewcommand{\theequation}{\thesection.\arabic{equation}}}

\newcommand{\dif}{\mathrm{d}}
\newcommand{\me}{\mathrm{e}}

\begin{document}
\addtolength{\baselineskip}{1.5mm}

\thispagestyle{empty}
\begin{flushright}

\end{flushright}
\vbox{}
\vspace{2cm}

\begin{center}
{\LARGE{Unbalanced Pomeransky--Sen'kov black ring
        }}\\[16mm]
{Yu Chen,~~Kenneth Hong~~and~~Edward Teo}
\\[6mm]
{\it Department of Physics,
National University of Singapore, 
Singapore 119260}\\[15mm]

\end{center}
\vspace{2cm}

\centerline{\bf Abstract}
\bigskip
\noindent
The Pomeransky--Sen'kov solution is well known to describe an asymptotically flat doubly rotating black ring in five dimensions, whose self-gravity is exactly balanced by the centrifugal force arising from the rotation in the ring direction. In this paper, we generalise this solution to the unbalanced case, in which there is in general a conical singularity in the space-time. Unlike a previous form of this solution presented in the literature, our form is much more compact. We describe in detail how this solution can be derived using the inverse-scattering method, and study its various properties. In particular, we show how various known limits can be recovered as special cases of this solution.


\newpage

\newsection{Introduction}

Although space-time appears to be four-dimensional, it has become apparent in recent years that a more complete understanding of general relativity can be obtained if the space-time dimensionality $D$ is made a tunable parameter. For instance, black holes in four dimensions are known to be subject to a number of uniqueness theorems, and it is of interest to see if these theorems are peculiar to four dimensions, or if they can be extended to higher dimensions. The Schwarzschild black-hole solution was first generalised to arbitrary dimension $D>4$ by Tangherlini in 1963 \cite{Tangherlini:1963bw}, while the rotating Kerr black-hole solution was similarly generalised by Myers and Perry in 1986 \cite{Myers:1986un}. Until a decade ago, these were the only higher-dimensional vacuum black holes known; in particular, it was still not clear then if there were uniqueness theorems to rule out other types of black holes in higher dimensions. 

In 2001, Emparan and Reall \cite{Emparan:2001wn} made a remarkable discovery of a new five-dimensional vacuum black hole with a non-spherical event-horizon topology. Instead, it has a ring topology $S^1\times S^2$, and was therefore called a black ring. This black ring rotates along the ring direction $S^1$, which creates a centrifugal force that opposes its self-gravity. Due to an imbalance of these two forces, there is a conical singularity in the space-time to stabilise the solution. However, for a certain value of the angular-momentum parameter, the forces balance exactly and there is no conical singularity present. The black ring is thus completely regular outside the event horizon. It turns out that, for a certain range of parameters, there are {\it two\/} regular black rings which share the same mass and angular momentum as the five-dimensional Myers--Perry black hole. This result unambiguously shows that the four-dimensional black-hole uniqueness theorems do not straightforwardly extend to higher dimensions.

Since black holes in five dimensions can rotate in two independent directions, it was natural to wonder if the Emparan--Reall black ring can be generalised to one with two independent rotations. A first step in this direction was made in 2005 by Mishima and Iguchi \cite{Mishima:2005id} and independently by Figueras \cite{Figueras:2005zp},\footnote{The solution was obtained in different coordinates in Ref.~\cite{Mishima:2005id} and \cite{Figueras:2005zp}.  In this paper, we shall use the Figueras form of the solution, which is the simpler of the two. The equivalence of the two forms was subsequently proved in \cite{Iguchi:2006tu}.} who discovered a solution describing a black ring that rotates only in the azimuthal direction of the $S^2$, i.e., there is no rotation along the ring direction. Because there is no centrifugal force in this case, the solution necessarily has a conical singularity to counteract the self-gravity of the black ring. The properties of this black ring were studied in detail in \cite{Iguchi:2006tu}.

It was by then clear that the most general doubly rotating black ring should contain both the above $S^1$-rotating and $S^2$-rotating black rings as special cases. It was also quite apparent that the form of this solution would be complicated, and that it could not be obtained by Wick-rotating a known solution (as was done in \cite{Emparan:2001wn}) or by ``educated guesswork'' (as was done in \cite{Figueras:2005zp}). A more systematic solution-generating technique was needed, and one that showed early promise was the inverse-scattering method (ISM) pioneered by Belinski and Zakharov \cite{Belinsky:1971,Belinsky:1979,Belinski:2001}. The usefulness of the ISM to higher-dimensional black holes was first pointed out by Pomeransky \cite{Pomeransky:2005sj}, who showed how to use it to obtain the five-dimensional Myers--Perry black hole by removing and adding solitons to a certain seed solution. Subsequently, it was shown in \cite{Tomizawa:2005wv} how the ISM could be used to generate the $S^2$-rotating black ring. However, using the ISM to generate the $S^1$-rotating black ring proved to be much subtler. The breakthrough in this came with the works of Iguchi and Mishima \cite{Iguchi:2006rd} and Tomizawa and Nozawa \cite{Tomizawa:2006vp}, who found the correct seed needed to generate the $S^1$-rotating black ring.

This progress paved the way for the generation of the doubly rotating black ring using the ISM. By combining the techniques used to generate the $S^1$-rotating and $S^2$-rotating black rings, this solution was first obtained by Pomeransky and Sen'kov \cite{Pomeransky:2006bd} in 2006. Although they mentioned that they had obtained the most general doubly rotating black ring solution, only the balanced case was presented in their paper. Furthermore, Pomeransky and Sen'kov found a form of the balanced doubly rotating black ring that was remarkably simple, considering the generality of the solution. The properties of this solution were further studied in \cite{Elvang:2007hs}. 

The unbalanced generalisation of the Pomeransky--Sen'kov black ring was eventually presented in \cite{Morisawa:2007di}. However, it took a very complicated form which made it difficult to handle and analyse. The main purpose of this paper is to present a much more compact form of this solution, which may be regarded as the natural generalisation of the simple form found in \cite{Pomeransky:2006bd} for the balanced case. We will also take the opportunity to describe in detail the ISM construction of this solution, something that was only briefly described in \cite{Pomeransky:2006bd} and not anywhere else to the best of our knowledge. 

We will also present a study of the physical properties of the unbalanced doubly rotating black ring. In particular, we show how the various known black-ring solutions, namely the Emparan--Reall, Figueras, and Pomeransky--Sen'kov solutions, can be obtained from it as special cases. We also explicitly show how the zero and infinite-ring radius limits of this solution can be taken, to obtain the general doubly rotating Myers--Perry black hole and general boosted Kerr black string, respectively. Note that the latter two limits would otherwise have been impossible to obtain from the Pomeransky--Sen'kov black ring. The fact that all these different limits can be obtained from our general solution is a good check that we have indeed found the correct unbalanced generalisation of the Pomeransky--Sen'kov black ring.

This paper is organised as follows: In Sec.~2, the ISM construction of the unbalanced doubly rotating black ring is described in detail. The final solution and its rod structure is presented in Sec.~3. The physical properties of this black ring are then discussed in Sec.~4. The paper ends with a brief discussion and an appendix containing several more technical results. Some familiarity with the ISM is assumed of the reader in Sec.~2. Those readers interested only in the solution and its physical properties may skip directly to Sec.~3.

\newsection{ISM construction}

The inverse-scattering method \cite{Belinsky:1971,Belinsky:1979,Belinski:2001} is a well-known technique that generates new solutions from known, simpler seed solutions by means of purely algebraic manipulations called soliton transformations. This, loosely speaking, refers to a paired process of removing and adding ``solitons'' to the seed solution, each with different ``Belinski--Zakharov vectors''. It is these BZ vectors which can be used to introduce new, non-trivial parameters to the seed solution. For example, the ISM can be used to introduce rotation to the Schwarzschild solution, thereby obtaining the Kerr solution. In recent years, it has been successfully applied to construct five-dimensional rotating black-hole solutions with $\mathbb{R}\times U(1)\times U(1)$ isometry (see, e.g., \cite{Pomeransky:2005sj,Emparan:2008eg,Iguchi:2011qi} for reviews). 

For such solutions, under suitable conditions, one can find so-called Weyl--Papapetrou coordinates in which their local metrics take the form \cite{Harmark:2004rm}
\be
\dif s^2=G_{ij}\dif x^i\dif x^j+\me^{2\gamma}(\dif \rho^2+\dif z^2)\,,
\ee
where the function $\gamma$ and the $3\times 3$ matrix $G_{ij}$ depend only on the two coordinates $\rho$ and $z$, with the latter satisfying the constraint $\det G=-\rho^2$. The seed solution will usually have a diagonal $G_{ij}$, and can be characterised by certain rod sources defined along the $z$-axis known as its rod structure \cite{Emparan:2001wk}. The new generated solution will in general have a non-diagonal $G_{ij}$. However, it can still be analysed using the rod-structure formalism developed in \cite{Harmark:2004rm,Hollands:2007aj,Chen:2010zu}. It turns out that the ISM and the rod-structure formalism together provide a powerful means to construct and analyse five-dimensional black holes with $\mathbb{R}\times U(1)\times U(1)$ isometry.

\begin{figure}[t]
\begin{center}
\includegraphics{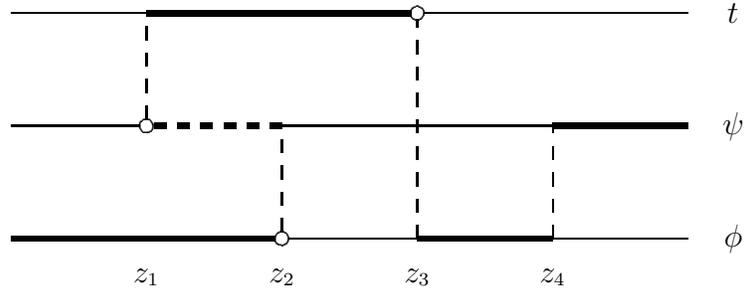}
\caption{The rod sources of the seed solution for the doubly rotating black ring. The thin lines denote the $z$-axis and the thick lines denote rod sources of mass $\frac{1}{2}$ per unit length along this axis. The dashed horizontal line denotes a rod source with negative mass density $-\frac{1}{2}$. Small circles represent the operations of removing solitons from the seed, each with a BZ vector having a single non-vanishing component along the coordinate that labels the $z$-axis where the circle is placed.}
\label{Seed for Pomeransky--Sen'kov black ring}
\end{center}
\end{figure}

In \cite{Pomeransky:2006bd}, Pomeransky and Sen'kov constructed their black ring in a two-step process. They first generated the Emparan--Reall black ring from a seed solution with the rod structure as shown in Fig.~\ref{Seed for Pomeransky--Sen'kov black ring}, using a one-soliton transformation in which a soliton was removed and added at $z_1$. A two-soliton transformation, in which solitons were removed and added at $z_2$ and $z_3$, was then performed on the balanced Emparan--Reall black ring to obtain the balanced doubly rotating black ring.

In this paper, we will use a three-soliton transformation on the same seed solution to directly generate the unbalanced doubly rotating black ring.\footnote{We remark that one can also start with the following seed solution:
\begin{equation*}
G_0={\textrm{diag}}\, \bigg\{ -{\frac {{\mu_1}}{{\mu_3}}},\frac{\mu_3\mu_4}{\mu_2}, \frac{\rho^2\mu_2}{\mu_1\mu_4}\bigg\}\,,
\end{equation*}
in which the rod with negative mass density lies between $z_2<z<z_3$, and generate the same final solution up to coordinate transformations and parameter redefinitions.} This has the advantage of being computationally simpler than the above two-step process. The explicit solution corresponding to the rod structure in Fig.~\ref{Seed for Pomeransky--Sen'kov black ring} can be directly read off \cite{Emparan:2001wk} as
\alpheqn
\ba
G_0&=&{\textrm{diag}}\, \bigg\{ -{\frac {{\mu_1}}{{\mu_3}}},\frac{\mu_2\mu_4}{\mu_1}, \frac{\rho^2\mu_3}{\mu_2\mu_4}\bigg\}\,,\\
\label{k}
\me^{2\gamma_0}&=&k^2\,\frac{\mu_2 \mu_4 R_{12} R_{13} R_{14} R_{23} R_{34}}{\mu_1 R_{24}^2 R_{11} R_{22} R_{33} R_{44}}\,,
\ea
\reseteqn
where $\mu_i\equiv\sqrt{\rho^2+(z-z_i)^2}-(z-z_i)$, $R_{ij}\equiv\rho^2+\mu_i\mu_j$, and $k$ is an arbitrary integration constant that can be fixed by, say requiring asymptotic flatness. An efficient calculation of the conformal factor $\me^{2\gamma_0}$ for a diagonal seed can be found in, say \cite{Izumi:2007qx}.
Using the ISM, we then perform the following soliton transformations on the above seed:
\begin{enumerate}
  \item Remove a soliton at each of $z_1$, $z_2$ and $z_3$, with trivial BZ vectors $(0,1,0)$, $(0,0,1)$ and $(1,0,0)$, respectively;
  \item Add back a soliton at each of $z_1$, $z_2$ and $z_3$, with non-trivial  BZ vectors $(C_1,1,0)$, $(0,C_2,1)$ and $(1,0,C_3)$, respectively. Here, $C_1$, $C_2$ and $C_3$ are the new, so-called BZ parameters. 
\end{enumerate}

In the first step, the act of removing a soliton at $z=z_k$ with a trivial BZ vector having a non-vanishing $a$-th component, refers to multiplying the diagonal element $(G_{0})_{aa}$ of the seed solution by a factor $-\frac{\mu_k^2}{\rho^2}$. So in the current case, after the first step, we obtain the new $G$-matrix:
\begin{equation}
\tilde{G}_0=G_0\,  {\textrm{diag}}\,  \bigg\{-\frac{\mu_3^2}{\rho^2},-\frac{\mu_1^2}{\rho^2},-\frac{\mu_2^2}{\rho^2}\bigg\}    = {\textrm{diag}}\,  \bigg\{\frac{\mu_1 \mu_3}{\rho^2} ,-{\frac {{\mu_1}{\mu_2}{\mu_4}}{{\rho}
^{2}}},-{\frac {{\mu_2}{\mu_3}}{{\mu_4}}} \bigg\}\,.
\end{equation}

To carry on with the second step, we need to know the corresponding $3\times3$ generating matrix $\tilde{\Psi}_0$. It can be obtained by performing the following replacements to the $\tilde{G}_0$ matrix: $\mu_i\rightarrow \mu_i-\lambda$, $\rho^2\rightarrow \rho^2-2z\lambda-\lambda^2$ or $\frac{\rho^2}{\mu_i}\rightarrow \frac{\rho^2}{\mu_i}+\lambda$, where $\lambda$ is a spectral parameter. Thus, we have
\be
\tilde{\Psi}_0(\lambda)={\textrm{diag}}\, \Bigg\{ \frac{(\mu_1-\lambda)( \mu_3-\lambda)}{(\mu_4-\lambda)\big(\frac{\rho^2}{\mu_4}+\lambda\big)},-{\frac {{(\mu_1-\lambda)}{(\mu_2-\lambda)}}{\frac{\rho^2}{\mu_4}+\lambda}},-{\frac {{(\mu_2-\lambda)}{(\mu_3-\lambda)}}{{\mu_4-\lambda}}} \Bigg\}\,.
\ee
One can then easily follow \cite{Pomeransky:2005sj} to carry out the second step. When adding back the solitons, we have to compute some vectors $m^{(k)}$ for the $k$-th soliton involving the quantity $\tilde\Psi_0^{-1}(\mu_k)$, which has infinite components. Such a difficulty can be circumvented by first multiplying $\tilde\Psi_0^{-1}(\lambda)$ by an overall factor $\lambda-\mu_k$ and then substituting $\lambda=\mu_k$ into the expressions. Although such an operation rescales $m^{(k)}$ by an infinite overall factor, it can be shown that it does not change the final solution. Note that after the first step, the matrix $\tilde{G}_0$ does not satisfy the constraint $\det \tilde{G}_0= -\rho^2$ as required of a physical solution; however, this constraint is automatically satisfied after we have done the second step.

To complete the construction, we need to calculate the conformal factor. It can be shown that the ratio of the new conformal factor to the old one is proportional to the determinant of the $\Gamma$-matrix as defined in \cite{Pomeransky:2005sj}, and it depends on the BZ parameters only through this determinant. Observe that in the above two steps the non-trivial BZ parameters $(C_1,C_2,C_3)$ only appear in the $\Gamma$-matrix of the second step. If we set $(C_1=0,C_2=0,C_3=0)$ so that the same solitons are first removed and then added back, we should be able to recover the original seed solution. It is then not difficult to see that
\be
\me^{2\gamma}=\me^{2\gamma_0}\frac{\det\Gamma(C_1,C_2,C_3)}{\det\Gamma(C_1=0,C_2=0,C_3=0)}\,,
\ee
where the $\Gamma$-matrix here is that corresponding to the second step.

At this stage, we have a new solution whose rod structure will be different from that in Fig.~\ref{Seed for Pomeransky--Sen'kov black ring}. Although there will still be four turning points and five rods, the directions of the rods will have components in all three possible directions, and will involve the BZ parameters $C_1$, $C_2$ and $C_3$ in a non-trivial way.
We now have to find the minimum conditions on $C_1$, $C_2$ and $C_3$ which would ensure that this new rod structure describes a black ring with the correct horizon topology \cite{Chen:2010ih}. Firstly, we require that the first and second rods (as counted from the left) are parallel, which gives an equation for $C_1{}^2$. Without loss of generality, we take the solution
\be
\label{C_1}
C_1=\sqrt {{\frac {z_{31}}{2z_{21}z_{41}}}}\,,
\ee
where $z_{ij}\equiv z_i-z_j$. This condition, in fact, implies another result: the orbits of the Killing vector fields associated with these two rods have the same periodicity. In the strengthened rod-structure formalism of \cite{Chen:2010zu}, these two rods then have the same {\it normalised\/} direction. Effectively, it means that the two rods are joined up into one, and that the point $z_1$ at which they meet is no longer a real turning point but a ``phantom point'' \cite{Emparan:2008eg}.

Secondly, we require the first and fourth rods to be parallel.\footnote{In general, these two rods will not have the same normalised direction. As we shall see below, the special case in which they do corresponds to the Pomeransky--Sen'kov black ring. In this case, $a^2-4b$ in (\ref{C3}) can be written as a perfect square, and $C_3$ becomes proportional to $C_2$. In this case, the subsequent analysis becomes much simpler.} This gives two conditions, one of which is actually guaranteed to hold by (\ref{C_1}). The second condition gives an equation quadratic in $C_3$. Of the two roots of this equation, we choose the one which contains as a special case the solution for the Pomeransky--Sen'kov black ring, namely
\be
\label{C3}
C_3=\sqrt {{\frac {z_{21}z_{42}{}^2}{ 2 z_{31} z_{41} }}}\,\frac{a-\sqrt{a^2-4b}}{C_2}\,,
\ee
where
\be
\label{ab}
a\equiv1+\frac{z_{32}}{z_{21}}\,{C_{2}}^{2},\qquad b\equiv\frac { z_{32}z_{41} z_{43}  }{
 z_{21} z_{42}{}^{2}}\,{C_{2}}^{2}.
\ee
The expression on the right-hand side of Eq.~(\ref{C3}) is rather unwieldy, and it would prove to be more convenient to define the new quantities $\alpha$ and $\beta$:
\be
\alpha=\frac{a+\sqrt{a^2-4b}}{2}\,,\qquad
\beta=\frac{a-\sqrt{a^2-4b}}{2}\,.
\ee
Then we can write $C_2$ and $C_3$ quite simply as
\be
\label{C2C3}
C_2=\sqrt{\frac{ z_{21} z_{42}{}^{2}\alpha\beta}{z_{32}z_{41} z_{43}}}\,,\qquad
C_3=\sqrt{\frac{2 z_{32} z_{43}\beta}{z_{31}\alpha}}\,.
\ee
Eq.~(\ref{ab}) will also imply an expression for the location of the phantom point, $z_1$: 
\be
\label{z1}
z_1=z_4-\frac{z_{42}{}^2}{z_{43}}\frac{\alpha\beta}{\alpha+\beta-1}\,,
\ee
in terms of $\alpha$, $\beta$ and the other $z_i$'s. 
Thus, with the use of (\ref{C2C3}) and (\ref{z1}), the solution can be expressed in terms of these parameters. Although there are apparently five of them, recall that the $z_i$ are only fixed up to an overall translation, so there are actually just four parameters which are physically relevant.

The next step involves rotating the solution to standard orientation \cite{Chen:2010zu} where the first and last space-like rods have directions $(0,0,1)$ and $(0,1,0)$, respectively. This will ensure that the metric takes a simple diagonal form at infinity, and is accomplished by the linear transformation $G'=A^TGA$, where
\be
A= \frac{1}{\zeta}\left( \begin {array}{ccc} \zeta~& K_5[1] & ~K_1[1] \\\noalign{\medskip}0~& K_5[2]& ~K_1[2]\\\noalign{\medskip}0~&K_5[3]& ~K_1[3]\end {array} \right).
\ee
Here, $K_1$ and $K_5$ are the {\it original\/} directions of the first and last space-like rods respectively, while $\zeta$ is a constant whose value is determined by the condition that $\det A=1$. 
These rod-directions (not normalised to have unit surface gravity) are given by
\ba
K_1&=&\left(2z_{32}z_{41}C_1C_2,~-z_{32}C_2,~z_{42}-z_{41} C_1C_2C_3\right),\cr
K_5&=&\left(z_{31}C_2C_3,~z_{42}-z_{41} C_1C_2C_3,~-z_{32}C_2\right).
\ea

Finally, we transform from Weyl--Papapetrou coordinates $(\rho,z)$ to C-metric-like coordinates $(x,y)$ \cite{Hong:2003gx}:
\ba
\label{C-metric}
\rho^2&=&\frac{4\varkappa^4(1-x^2)(y^2-1)(1+\mu x)(1+\mu y)(1+\nu x)(1+\nu y)}{(1-\mu\nu)^2(x-y)^4}\,,\cr
z&=&\frac{\varkappa^2(1-xy)\left[2+(\mu+\nu)(x+y)+2\mu\nu xy\right]}{(1-\mu\nu)(x-y)^2}\,,
\ea
with the locations of the three remaining turning points fixed to be
\be
z_2=-\frac{\mu-\nu}{1-\mu\nu}\,\varkappa^2,\qquad
z_3=\frac{\mu-\nu}{1-\mu\nu}\,\varkappa^2,\qquad
z_4=\varkappa^2.
\ee
The key advantage of using C-metric-like coordinates is that $\mu_i$, $i=2,3,4$ become algebraic expressions:
\ba
\mu_2&=&-\frac{2\varkappa^2(1-x)(1+y)(1+\nu x)(1+\mu y)}{(1-\mu\nu)(x-y)^2}\,,\cr
\mu_3&=&-\frac{2\varkappa^2(1-x)(1+y)(1+\mu x)(1+\nu y)}{(1-\mu\nu)(x-y)^2}\,,\cr
\mu_4&=&-\frac{2\varkappa^2(1-y^2)(1+\mu x)(1+\nu x)}{(1-\mu\nu)(x-y)^2}\,.
\ea
Note that this particular coordinate transformation introduces one new parameter, and the additional freedom can be used to simplify the metric components of the resulting solution. Following the case of the Pomeransky--Sen'kov black ring \cite{Pomeransky:2006bd}, we use this freedom to make the numerator of $g_{t\phi}$ linear in $y$. This fixes $\beta$ as follows:
\be
\label{beta}
\beta=\frac { 1-\mu }{2\mu}  \left( \frac{ \nu-\mu }{1+\nu}\, \alpha+ \frac{ \mu+\nu }{1-\nu} \right).
\ee
Furthermore, it is observed that the metric components take the simplest form when $\alpha$ is eliminated in favour of the new parameter $\lambda$ (not to be confused with the spectral parameter used above) as follows:
\be
\label{alpha}
\alpha={\frac { \left( 1+\nu \right) \left( \mu-\lambda\nu+\mu\nu-\lambda\mu^2\right) }{ \left( 1-\nu \right) \left( \mu+\lambda\nu-\mu\nu-\lambda\mu^2\right) }}\,.
\ee
After making a suitable choice of the integration constant $k$ in (\ref{k}) to ensure asymptotic flatness (and a possible sign change $t\rightarrow-t$), the metric of the resulting solution is given below in Eq.~(\ref{general_BR}) in terms of the final physical parameters $\mu$, $\nu$, $\lambda$, and $\varkappa$.

We remark that while the above-described procedure involves only straightforward algebraic manipulations from start to finish, the expressions involved can become very complicated in the intermediate stages, making them a challenge to handle even on modern computers. This is because they will involve the explicit square-root $\sqrt{\rho^2+(z-z_1)^2}$ coming from $\mu_1$, which cannot be avoided even after transforming to the C-metric-like coordinates (\ref{C-metric}).
Although we know that the final metric will not depend on this square-root (since $z_1$ will turn into a phantom point upon imposing the condition (\ref{C_1})), the challenge is to try to cancel out the square-root at some stage.\footnote{This step also needs to be carried out in the ISM generation of the Emparan--Reall black ring, and is described, for example, in Appendix A.2 of \cite{Elvang:2007rd}.} It is obviously desirable to do this as early in the computation as possible.

In practice, we have found that the following (supplemented) sequence of steps will enable the whole computation to be done on a modern computer in reasonable time: The three-soliton transformation is first carried out in Weyl--Papapetrou coordinates. We then transformed the resulting solution to the C-metric-like coordinates $(u,v)$ described in Appendix H of \cite{Harmark:2004rm}. These C-metric-like coordinates contain one fewer parameter than those in (\ref{C-metric}), but they have the advantage of being simpler in form. In what turns out to be the most computationally intensive step, we then imposed the condition (\ref{C_1}) and cancelled out all the square-roots $\sqrt{\rho^2+(z-z_1)^2}$ using a similar procedure to that described in \cite{Elvang:2007rd}. After this is done, the remaining conditions (\ref{C2C3}) and (\ref{z1}) are imposed, and the rotation to standard orientation performed. Finally, we performed the following M\"obius transformation:
\be
u=\frac{x+\nu}{1+\nu x}\,,\qquad v=\frac{y+\nu}{1+\nu y}\,,\qquad
c=\frac{\mu-\nu}{1-\mu\nu}\,,
\ee
to transform from the C-metric-like coordinates of \cite{Harmark:2004rm} to those in (\ref{C-metric}), and the expressions (\ref{beta}) and (\ref{alpha}) are substituted in to obtain the final form of the solution.

\newsection{The metric and rod structure}

The metric of the unbalanced doubly rotating black ring is given by
\ba
\label{general_BR}
\dif s^2&=&-\frac{H(y,x)}{H(x,y)}\,\left(\dif t-\omega_\psi\,\dif\psi-\omega_\phi\,\dif\phi\right)^2-\frac{F(x,y)}{H(y,x)}\,\dif\psi^2-2\,\frac{J(x,y)}{H(y,x)}\,\dif\psi\,\dif\phi+\frac{F(y,x)}{H(y,x)}\,\dif\phi^2\nonumber\\
&&{}+\frac{2\varkappa^2(1-\mu)^2(1-\nu)H(x,y)}{(1-\lambda)(1-\mu\nu)\Phi\Psi(x-y)^2}\left[\frac{\dif x^2}{G(x)}-\frac{\dif y^2}{G(y)}\right],
\ea
where
\ba
\omega_\psi&=&\frac{\varkappa(\mu+\nu)}{H(y,x)}\,\sqrt{\frac{2\lambda(\lambda-\mu)(1+\lambda)(1-\lambda\mu)\Phi\Xi}{(1-\lambda)(1-\mu\nu)\Psi}}\,(1+y)\cr
&&\times\left\{\Phi ( 1+\nu{x}^{2}y ) +\nu(1-\mu)\left[1+\lambda x -xy(x+\lambda)\right]\right\},\cr
\omega_\phi&=&\frac{\varkappa(\mu+\nu)}{H(y,x)}\,\sqrt{\frac{2\nu\lambda(1-\lambda^2)\Phi\Psi\Xi}{1-\mu\nu}}\,(1-x^2)y\,,
\ea
and the functions $G$, $H$, $J$ and $F$ are given by
\alpheqn
\ba
G(x)&=&(1-x^2)(1+\mu x)(1+\nu x)\,,\\
H(x,y)&=&\Phi\Psi+\nu(\lambda-\mu)(1+\lambda)\Phi
+\nu\Psi\Xi{x}^{2}{y}^{2}+\nu ( \mu+\nu )(\lambda-\mu )(1- \lambda\mu)( 1-\lambda\mu{x}^{2}{y}^{2} ) \cr
&& +\lambda ( \mu+\nu
 )  [ 1-\lambda\mu-\nu (\lambda-\mu)xy
 ]  [  ( 1-\lambda\mu ) x+ \nu( \lambda-\mu
 ) y ],\\
J(x,y)&=&\frac{2\varkappa^2(\mu+\nu)\,\sqrt{\nu(\lambda-\mu)(1-\lambda\mu)}(1-x^2)(1-y^2)}{(1-\mu\nu)\Phi(x-y)}\,\Big\{
\Phi\Psi+\nu(\lambda-\mu)(1+\lambda)\Phi\cr
&&-\nu\Psi\Xi xy+\nu  
 ( \mu+\nu )( \lambda-\mu ) ( 1-\lambda\mu ) ( 1+\lambda x+\lambda y+\lambda\mu xy)\Big\}\,,\\
F(x,y)&=&\frac{2\varkappa^2}{\mu\nu(1-\mu\nu)\Phi(x-y)^2}\,\Bigg\{G(x)(y^2-1)\,\bigg\{\mu(1-\lambda^2)[\Psi+\nu(\lambda-\mu)(1+\nu)]^2\cr
&&-(\mu+\nu)(1-\lambda\mu)(1+\nu y)\Big[\Psi\Xi-\lambda\mu(\lambda-\mu)[\Psi+\nu(\lambda-\mu)(1+\nu)]\Big]\bigg\}\cr
&&+\nu G(y)\bigg\{(\lambda-\mu)(1-\lambda\mu)\Big[
\lambda(\mu+\nu)^2(1-\lambda\mu)+[\Psi+\nu(\lambda-\mu)(1+\nu)]\cr
&&\times(\mu+\nu-\mu\nu x)x\Big]+[\Psi\Xi+\lambda\mu\Phi(\Phi-1)(\Phi-\Psi+\Xi) 
][1+(\mu+\nu)x]x^2\cr
&&+\mu\nu\Phi[\Psi\Xi-\lambda\mu(\mu+\nu)(\lambda-\mu)(1-\lambda\mu)]x^4
\bigg\}\Bigg\}\,.
\ea
\reseteqn
To simplify the above expressions, we have introduced the following abbreviations:
\be
\Phi\equiv1-\lambda\mu-\lambda\nu+\mu\nu\,,\qquad
\Psi\equiv\mu-\lambda\nu+\mu\nu-\lambda\mu^2\,,\qquad
\Xi\equiv\mu+\lambda\nu-\mu\nu-\lambda\mu^2\,.
\ee
The C-metric-like coordinates $x$, $y$ take the ranges $-1\le x\le 1$ and $-\infty<y\le-1$, respectively. The metric is independent of time $-\infty<t<\infty$ and angles $0\le\psi,\phi<2\pi$. It has four independent physical parameters, $\mu$, $\nu$, $\lambda$ and $\varkappa$, of which the first three are dimensionless and the last sets the scale of the solution. They satisfy the constraints
\be
\label{param_ranges}
0\le\nu\le\mu\le\lambda<1\,,\qquad\varkappa>0\,.
\ee
In particular, the former constraint ensures that the quantities $\Phi$, $\Psi$ and $\Xi$ are positive.

\begin{figure}[t]
\begin{center}
\includegraphics{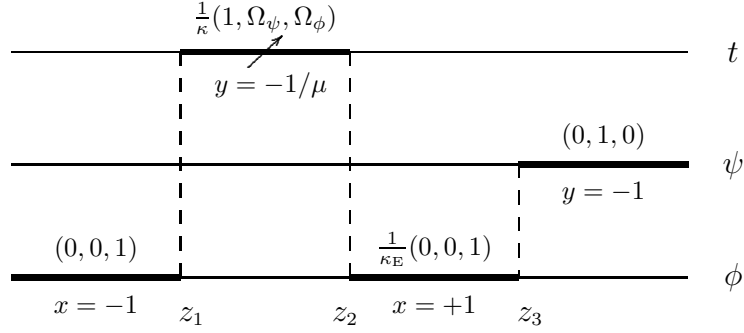}
\caption{The rod structure of the unbalanced doubly rotating black ring. The location of each rod is indicated below it, while the direction of each rod is indicated above it. The arrow on the horizon rod indicates that its direction vector has components in the $\psi$ and $\phi$ directions as well.}
\label{Final rod structure}
\end{center}
\end{figure}

As a check, we can calculate the rod structure of this solution using standard methods \cite{Chen:2010zu,Chen:2010ih}. It indeed has three turning points, consistent with the fact that the ISM construction above has turned one of the original four turning points into a phantom point. They are located at\footnote{There is no longer any need to enumerate the phantom point, since its position is determined in terms of the other parameters of the solution.} $(\rho=0,z=z_1\equiv-\frac{\mu-\nu}{1-\mu\nu}\,\varkappa^2)$ or $(x=-1,y=-1/\mu)$, $(\rho=0,z=z_2\equiv\frac{\mu-\nu}{1-\mu\nu}\,\varkappa^2)$ or $(x=1,y=-1/\mu)$, and $(\rho=0,z=z_3\equiv\varkappa^2)$ or $(x=1,y=-1)$, respectively. These three turning points partition the $z$-axis into four rods; from left to right they are:
\begin{itemize}
\item Rod 1:\enskip a semi-infinite space-like rod at $(x=-1, -1/\mu\le y<-1)$, with direction $\ell_1=(0,0,1)$.
\item Rod 2:\enskip a finite time-like rod at $(-1\le x\le 1, y=-1/\mu)$, with direction $\ell_2=\frac{1}{\kappa}(1,\Omega_\psi,\Omega_\phi)$, where 
\alpheqn
\ba
\label{kappa}
\kappa&=&\frac{(\mu-\nu)(1-\lambda)(1+\mu)}{4\varkappa(\mu+\nu)(1-\mu)\Xi}\,\sqrt{\frac{2(1-\mu\nu)\Phi\Psi}{\lambda(1+\lambda)(1-\nu)}}\,,\\
\label{Omega_psi}
\Omega_\psi&=&\frac{1}{\varkappa(1-\mu)}\,\sqrt{\frac{(\lambda-\mu)(1-\lambda)(1-\lambda\mu)(1-\mu\nu)\Psi}{2\lambda(1+\lambda)\Phi\Xi}}\,,\\
\label{Omega_phi}
\Omega_\phi&=&\frac{1+\mu}{\varkappa(\mu+\nu)}\,\sqrt{\frac{\nu(1-\lambda)(1-\mu\nu)\Psi}{2\lambda(1+\lambda)\Phi\Xi}}\,.
\ea
\reseteqn
\item Rod 3:\enskip a finite space-like rod at $(x=1, -1/\mu\le y\le-1)$, with direction $\ell_3=\frac{1}{\kappa_\mathrm{E}}(0,0,1)$, where 
\be
\label{kappa_E}
\kappa_{\mathrm{E}}=\frac{1+\mu}{1-\mu}\,\sqrt{\frac{(1-\lambda)(1+\nu)\Psi}{(1+\lambda)(1-\nu)\Xi}}\,.
\ee
\item Rod 4:\enskip a semi-infinite space-like rod at $(-1<x\le1, y=-1)$, with direction $\ell_4=(0,1,0)$.
\end{itemize}
This rod structure is shown schematically in Fig.~\ref{Final rod structure}, and much information can already be read off from it using results from the rod-structure formalism \cite{Chen:2010zu,Chen:2010ih}. It is clear that it describes a doubly rotating black ring in an asymptotically flat space-time. In the following section, we shall examine the physical properties of this black ring in more detail.

\newsection{Physical properties}

We begin by noting that the metric of the unbalanced doubly rotating black ring (\ref{general_BR}) is asymptotically flat, with infinity located at $(x,y)\rightarrow(-1,-1)$. This can be explicitly seen by introducing the coordinates $(r,\theta)$ defined by
\be
x=-1+\frac{4\varkappa^2(1-\mu)(1-\nu)}{1-\mu\nu}\,\frac{\cos^2\theta}{r^2}\,,\qquad
y=-1-\frac{4\varkappa^2(1-\mu)(1-\nu)}{1-\mu\nu}\,\frac{\sin^2\theta}{r^2}\,.
\ee
In the asymptotic region $r\to\infty$, the metric behaves as
\ba
\dif s^2&\rightarrow&\left(-1+\frac{8M}{3\pi r^2}\right)\dif t^2-\frac{8J_\psi\sin^2\theta}{\pi r^2}\,\dif t\,\dif\psi-\frac{8J_\phi\cos^2\theta}{\pi r^2}\,\dif t\,\dif\phi\nonumber\\
&&{}+\dif r^2+r^2\left(\dif\theta^2+\sin^2\theta\,\dif\psi^2+\cos^2\theta\,\dif\phi^2\right),
\ea
from which we can read off the ADM mass $M$ and angular momenta $J_\psi$, $J_\phi$ of the space-time:
\alpheqn
\ba
M&=&\frac{3\pi\varkappa^2\lambda(\mu+\nu)(1-\mu)\Phi}{2(1-\lambda)(1-\mu\nu)\Psi}\,,\\
J_\psi&=&\frac{\pi\varkappa^3(\mu+\nu)(1-\mu)[2\nu(1-\lambda)(1-\mu)+(1-\nu)\Phi]}{(1-\lambda)^{3/2}(1-\mu\nu)^{3/2}\Psi^{3/2}}\cr
&&\times\sqrt{\frac{2\lambda(\lambda-\mu)(1+\lambda)(1-\lambda\mu)\Xi}{\Phi}}\,,\\
J_\phi&=&\frac{2\pi\varkappa^3(\mu+\nu)(1-\mu)}{(1-\mu\nu)^{3/2}}\,\sqrt{\frac{2\nu\lambda(1+\lambda)\Xi}{(1-\lambda)\Phi\Psi}}\,.
\ea
\reseteqn
Although these three physical quantities depend on the four parameters in a rather non-trivial way, it is possible to read off some special cases bearing in mind the parameter ranges (\ref{param_ranges}). $J_\psi$ can be made to vanish by setting $\lambda=\mu$, while $J_\phi$ can be made to vanish by setting $\nu=0$. When both angular momenta vanish, (\ref{general_BR}) reduces to the static black ring \cite{Emparan:2001wk}. All three quantities $M$, $J_\psi$ and $J_\phi$ can be made to vanish by setting $\nu=\mu=\lambda=0$, in which case it can be checked that (\ref{general_BR}) reduces to flat Minkowski space-time.

As can be seen from the rod structure, there is an event horizon located at $y=-1/\mu$. It has a ring topology $S^1\times S^2$, with $\partial/\partial\psi$ generating the $S^1$ and $\partial/\partial\phi$ generating the rotational symmetry of the $S^2$. The surface gravity $\kappa$, and angular velocities $\Omega_\psi$ and $\Omega_\phi$, of the event horizon are given by (\ref{kappa}), (\ref{Omega_psi}) and (\ref{Omega_phi}), respectively. Its area can be computed to be\footnote{If we identify the temperature and entropy of the event horizon as $T=\kappa/(2\pi)$ and $S=A/4$ respectively, it can be checked that the Smarr relation
\begin{equation*}
\frac{2}{3}M=TS+\Omega_\psi J_\psi+\Omega_\phi J_\phi\,,
\end{equation*}
is satisfied.} 
\be
A=\frac{16\pi^2\varkappa^3(\mu+\nu)(1-\mu)\Xi}{(1-\lambda)(1+\mu)(1-\mu\nu)^{3/2}}\,\sqrt{\frac{2\lambda(1+\lambda)(1-\nu)}{\Phi\Psi}}\,.\\
\ee
Although the present coordinates break down at the horizon, it is possible to find good coordinates through it, following the methods of, say \cite{Elvang:2007hs,Durkee:2008an}. It turns out that there is an inner horizon at $y=-1/\nu$, and a curvature singularity beyond it where $H(x,y)=0$. For the most part, we will concentrate on the region outside the event horizon, namely $-1/\mu<y\leq-1$. 

Note that $g_{t\phi}$ vanishes along the two axes $x=\pm1$. This ensures that Dirac--Misner singularities are absent in the space-time. The absence of such singularities can in fact be seen from the rod structure, from the fact that the directions of Rods 1 and 3 do not have time-like components \cite{Chen:2010zu,Chen:2010ih}. It can also be checked that $H(x,y)>0$ everywhere on and outside the horizon; see Appendix A. Since the curvature invariants have denominators that are proportional to some positive power of $H(x,y)$, there are no curvature singularities on or outside the horizon. 

However, in general, there will be a conical singularity present in the space-time, located along the finite axis $x=1$. From the rod-structure viewpoint, this can be seen from the direction vector of Rod 3, which is not canonically normalised to $(0,0,1)$ like that of Rod 1. There is a conical excess along this axis, which can be read off to be
\be
\label{Delta_phi}
\Delta\phi=2\pi(\kappa_\mathrm{E}-1)\,,
\ee
where $\kappa_\mathrm{E}$ is given by (\ref{kappa_E}). This conical (strut) singularity provides a pressure that prevents the black ring from collapsing under its own gravity. However, when the rotation in the ring direction is sufficiently large, the conical excess in (\ref{Delta_phi}) becomes a deficit, and the conical singularity provides a tension that prevents the black ring from breaking apart due to the centrifugal force.

It is clear by now that (\ref{general_BR}) describes a black ring which has rotations along two independent directions, with a conical singularity in the space-time.  This physical picture can be confirmed by seeing how some well-known solutions can be recovered from (\ref{general_BR}) as special cases. The first limit we shall consider is when there is {\it no\/} conical singularity in the space-time. This is given by the condition $\kappa_\mathrm{E}=1$, which gives two solutions when solved for $\lambda$. It turns out that only one of them,
\be
\label{regularity_condition}
\lambda=\frac{2\mu}{1+\mu^2}\,,
\ee
lies within the physical range given by (\ref{param_ranges}).
This limit should correspond to the Pomeransky--Sen'kov black ring \cite{Pomeransky:2006bd}, and indeed we can recover this solution from (\ref{general_BR}) upon imposing the condition (\ref{regularity_condition}). This is explicitly shown in Appendix B (together with the other well-known limits to be discussed below).

If we set $\nu=0$, we recover from (\ref{general_BR}) the Emparan--Reall black ring \cite{Emparan:2001wn}, which rotates only in the $\psi$ direction. On the other hand, if we set $\lambda=\mu$, we recover the Figueras black ring \cite{Figueras:2005zp}, which rotates only in the $\phi$ direction.
These two results are consistent with the observation made above that $J_\phi=0$ and $J_\psi=0$ in these two limits, respectively. Roughly speaking, they also show that the parameter $\nu$ governs the rotation of the black ring in the $S^2$ direction, while the parameter $\lambda$ governs its rotation in the $S^1$ direction. In particular, the upper bound for $\nu$, namely $\nu=\mu$, describes a black ring that is maximally rotating in the $S^2$ direction. This corresponds to the extremal limit of (\ref{general_BR}) in which the two horizons coincide, and their surface gravity becomes zero.

There are two other non-trivial limits of (\ref{general_BR}) that one can check, namely the zero and infinite ring-radius limits. As expected, when the ring-radius is shrunk to zero, the black ring becomes a Myers--Perry black hole that rotates in two independent directions. On the other hand, when the ring-radius becomes infinite, the $S^1$ decompactifies into an $\mathbb{R}^1$ and the black ring becomes an infinitely extended black string. The original $S^1$-rotation now corresponds to momentum in the string direction, while the $S^2$-rotation is orthogonal to the string direction. In other words, we have a boosted Kerr black string. This conclusion is consistent with the general interpretation of black rings as black strings that are bent into a circular shape.

Since rotation is present, there will be an ergoregion in the space-time bounded by ergosurfaces on which $H(y,x)=0$. It turns out that the properties of this ergoregion are qualitatively similar to that of the Pomeransky--Sen'kov black ring \cite{Elvang:2008qi,Durkee:2008an}. In particular, it can be checked that the event horizon is always surrounded by the ergoregion. For sufficiently small $\nu$ (and fixed $\lambda$ and $\mu$), the topology of the ergosurface is $S^1\times S^2$, just like that of the event horizon. However, there exists a critical value of $\nu$, given by
\be
\label{merger_cond}
\nu=\frac{1-\lambda\mu}{1+\lambda}\,,
\ee
beyond which the ergoregion will grow large enough to merge with itself across the centre of the ring. At the same time, an inner $S^3$ ergosurface will appear, so as to exclude the centre of the ring from the ergoregion. The ergoregion will then encompass the region between the outer and inner $S^3$ ergosurfaces; see, e.g., Fig.~1 of \cite{Durkee:2008an}. This means the topology of the ergosurface will change from $S^1\times S^2$ to $S^3\cup S^3$. It should be noted that the condition (\ref{merger_cond}) is not always compatible with the physical range (\ref{param_ranges}), i.e., there exist values of $\lambda$ and $\mu$ for which the topology of the ergosurface is {\it always\/} $S^1\times S^2$.

It should also be checked if the space-time described by (\ref{general_BR}) contains closed time-like curves (CTCs). Now, the requirement for the absence of CTCs is that the $2\times2$ metric $g_{ij}$, $i,j=\psi,\phi$ be positive semi-definite. Since the metric components are sufficiently complicated, we have resorted to checking this numerically. Despite an extensive search, no CTCs were found anywhere in the space-time outside the event horizon. It would be desirable to demonstrate the absence of CTCs analytically, perhaps along the lines of \cite{Chrusciel:2010jg} as done for the Pomeransky--Sen'kov black ring.

\newsection{Discussion}

In this paper, we have described how the inverse-scattering method can be used to derive the unbalanced doubly rotating black ring, which generalises the Pomeransky--Sen'kov black ring \cite{Pomeransky:2006bd}. We then presented a new form of this solution which is much more compact than the one previously presented in \cite{Morisawa:2007di}. Finally, we studied some physical properties of this solution, including showing how various well-known limits can be obtained from it.

There are several possible extensions of this work. While we have studied the main properties of the unbalanced doubly rotating black ring, there are other properties that are worth investigating in more detail, such as its geodesics and global structure (as was done in \cite{Durkee:2008an} and \cite{Chrusciel:2009vr,Chrusciel:2010jg}, respectively, for the balanced case). For completeness, it would also be desirable to map the form of the solution found in this paper to the one presented in \cite{Morisawa:2007di}.

It should be possible to generalise the black-ring solution found in this paper to include charge. Unlike black holes, black rings can carry two kinds of charge: a normal conserved charge \cite{Elvang:2003yy} and a non-conserved dipole charge \cite{Emparan:2004wy}. It should be straightforward to obtain charged doubly rotating black rings using standard charging transformations, such as those used in \cite{Elvang:2003yy,Bouchareb:2007ax,Hoskisson:2008qq,Gal'tsov:2009da}. However, obtaining a dipole-charged doubly rotating black ring might prove to be more elusive. At present, even the dipole-charged generalisation of the $S^2$-rotating black ring is not known, although the inverse-scattering formalism developed in \cite{Figueras:2009mc} might offer some hope for finding this solution. Such a solution, if found, could be the starting point to generate the most general stationary black-ring solution of $U(1)^3$ supergravity theory \cite{Elvang:2004xi}.

It is our hope that the ISM construction explicitly presented in this paper will inspire the construction of other more general solutions, such as those containing multiple black rings/holes in five dimensions, with rotations in two independent directions. An example is a doubly rotating black ring with a Myers--Perry black hole at its centre, a configuration known as a black saturn \cite{Elvang:2007rd}. Other possible configurations consist of two concentric doubly rotating black rings lying in the same plane (known as a di-ring \cite{Iguchi:2007is,Evslin:2007fv}), or in orthogonal planes (known as a bi-ring \cite{Izumi:2007qx,Elvang:2007hs}). Each of these solutions is expected to have a regular sub-class of solutions, which could have implications, for example, for the phase structure of black holes in five dimensions \cite{Elvang:2007hs}.

Finally, we remark that the ISM construction of the doubly rotating black ring can be extended to obtain a doubly rotating ``black lens'' with a lens-space horizon topology, generalising the singly rotating black lens found in \cite{Chen:2008fa}. The ISM construction in this case is rather straightforward: we use the same seed solution, with a rod structure as shown in Fig.~1, but now we need to remove a fourth soliton at $z_4$ with a trivial BZ vector $(0,0,1)$ in Step 1 above, and then add it back with a non-trivial BZ vector $(0,C_4,1)$ in Step 2 together with the other three solitons. After imposing suitable conditions \cite{Chen:2008fa}, we obtain a black lens rotating in two independent directions. We also mention that the ISM construction presented in this paper can be extended to obtain a doubly rotating black ring on Taub-NUT, generalising the black ring on Taub-NUT with a single angular-momentum parameter found in \cite{Camps:2008hb}.

\bigbreak\bigskip\bigskip\centerline{{\bf Acknowledgement}}
\nobreak\noindent This work was partially supported by the Academic Research Fund (WBS No.: R-144-000-277-112) from the National University of Singapore.

\appendix

\newsection{Positivity of $H(x,y)$}

Here, we prove that $H(x,y)$ is positive everywhere on and outside the event horizon $-1\leq x\leq1$, $-1/\mu\leq y\leq-1$, and for the range of parameters\footnote{We exclude the Emparan--Reall black-ring limit here, as $H(x,y)$ is already known to be positive in this case.} $0<\nu\le\mu\le\lambda<1$. We begin by noting that $H(x,y)$ is a quadratic function in $x$ that can be written in the form:
\be
H(x,y)=ax^2+bx+c\,,
\ee
where $a$, $b$ and $c$ are functions of $y$, given by
\ba
a&=&\nu y[\Psi\Xi y-\lambda(1-\lambda\mu)(\lambda-\mu)(\mu+\nu)(1+\mu y)]\,,\cr
b&=&\lambda( \mu+\nu ) [ ( 1-\lambda\mu ) ^{2}-
 {\nu}^{2} ( \lambda-\mu )^{2}{y}^{2}]\,,\cr
c&=&\Phi[ \Psi+\nu(\lambda-\mu)(1+\lambda)] +\nu( 1-\lambda\mu)  ( \lambda-\mu) ( \mu+\nu )  ( 1+\lambda y )\,.
\ea
$a$ is manifestly positive for the above-stated physical range. For later purposes, we note that $b$ is also positive in this range:
\be
b\geq\lambda( \mu+\nu ) \bigg[ ( 1-\lambda\mu ) ^{2}-
 {\nu}^{2} ( \lambda-\mu )^{2}\bigg(-\frac{1}{\mu}\bigg)^{2}\bigg]=\frac{1}{\mu^2}\lambda(\mu+\nu)\Psi\Xi>0\,.
\ee
It turns out that $c$ is positive in this range as well, but we shall show the stronger result that $c\geq a$. This can be seen from the fact that
\be
\frac{\dif}{\dif y}(c-a)=2\nu[-\Psi\Xi y+\lambda(1-\lambda\mu)(\lambda-\mu)(\mu+\nu)(1+\mu y)]>0\,,
\ee
and
\be
(c-a)\Big|_{y=-1/\mu}=\frac{1}{\mu^2}(\mu-\nu)\Psi\Xi\geq0\,.
\ee

Now, assume that $b^2-4ac\geq0$. (Otherwise $H(x,y)$ will not have any zeros, which would mean it is positive everywhere since $a>0$.) The minimum of $H(x,y)$ with respect to $x$ is located at
\be
x_0\equiv-\frac{b}{2a}\,.
\ee
To prove that $H(x,y)$ is positive in the range $-1\leq x\leq1$, we need to show that (i) $H(-1,y)>0$, and (ii) $x_0\leq-1$. 
It is readily seen that the first condition is satisfied:
\ba
H(-1,y)&=&\frac {1}{\mu} ( 1-\lambda )\Psi \big[  ( 1+\mu\nu{y}^{2} ) \Xi+\lambda\nu ( 1+\mu ) ( {
\mu}^{2}{y}^{2}-1)  \big]\cr
&\geq&\frac {1}{\mu} ( 1-\lambda )\Psi \big[  ( 1+\mu\nu ) \Xi+\lambda\nu ( 1+\mu ) ( {
\mu}^{2}-1)  \big]\cr
&=&(1-\lambda)(1-\nu)\Phi\Psi>0\,.
\ea
The second condition can be shown as follows: since $b>0$ and $c\geq a>0$, it follows from the initial assumption that $b\geq2\sqrt{ac}\geq2a$, and so $-\frac{b}{2a}\leq-1$. 
Hence $H(x,y)$ is positive on and outside the event horizon.

\newsection{Various known limits}

\newsubsection{Pomeransky--Sen'kov black ring}

The Pomeransky--Sen'kov black ring \cite{Pomeransky:2006bd} is obtained by imposing the regularity condition along the finite axis, namely (\ref{regularity_condition}). The metric (\ref{general_BR}) then becomes
\ba
\label{Pomeransky_BR}
\dif s^2&=&-\frac{\tilde{H}(y,x)}{\tilde{H}(x,y)}\,\left(\dif t-\tilde{\omega}_\psi\,\dif\psi-\tilde{\omega}_\phi\,\dif\phi\right)^2-\frac{\tilde{F}(x,y)}{\tilde{H}(y,x)}\,\dif\psi^2-2\,\frac{\tilde{J}(x,y)}{\tilde{H}(y,x)}\,\dif\psi\,\dif\phi+\frac{\tilde{F}(y,x)}{\tilde{H}(y,x)}\,\dif\phi^2\nonumber\\
&&{}+\frac{2\varkappa^2\tilde{H}(x,y)}{(1-\mu\nu)^2(x-y)^2}\left[\frac{\dif x^2}{G(x)}-\frac{\dif y^2}{G(y)}\right],
\ea
where
\ba
\tilde{\omega}_{\psi}&=&\frac{2\varkappa(\mu+\nu)}{\tilde H(y,x)}\sqrt{\frac{(1+\mu)(1+\nu)}{(1-\mu)(1-\nu)}}\,(1+y)\cr
&&\times\left[1+\mu+\nu-\mu\nu+2\mu\nu x(1-y)+\mu\nu(1-\mu-\nu-\mu\nu)x^2y\right],\cr
\tilde{\omega}_{\phi}&=&\frac{2\varkappa(\mu+\nu)}{\tilde H(y,x)}\sqrt{\mu\nu(1-\mu^2)(1-\nu^2)}\,(1-x^2)y\,,
\ea
and the functions $\tilde H$, $\tilde J$ and $\tilde F$ are given by
\alpheqn
\ba
\tilde{H}(x,y)&=&1+(\mu+\nu)^2-\mu^2\nu^2+2(\mu+\nu)(x+\mu\nu y)(1-\mu\nu xy)\nonumber\\
&&{}+\mu\nu\big[1-(\mu+\nu)^2-\mu^2\nu^2\big]x^2y^2\,, \\
\tilde{J}(x,y)&=&\frac{2\varkappa^2(\mu+\nu)\sqrt{\mu\nu}(1-x^2)(1-y^2)}{(1-\mu\nu)^2(x-y)}\,\Big\{1+(\mu+\nu)^2-\mu^2\nu^2\cr
&&{}-\mu\nu\big[1-(\mu+\nu)^2-\mu^2\nu^2\big]xy+2\mu\nu(\mu+\nu)(x+y)\Big\}\,, \\
\tilde{F}(x,y)&=&\frac{2\varkappa^2}{(1-\mu\nu)^2(x-y)^2}\,\Bigg\{G(x)(1-y^2)\Big\{(1+\mu\nu)\big[1-(\mu+\nu)^2-2\mu\nu+\mu^2\nu^2\big]\nonumber\\
&&{}+(\mu+\nu)(1-\mu^2-\nu^2-3\mu^2\nu^2)y\Big\}+G(y)\Big\{2(\mu+\nu)^2+(\mu+\nu)(1+\mu^2)\nonumber\\
&&{}\times(1+\nu^2)x+(1+\mu\nu)\big[1-(\mu+\nu)^2-2\mu\nu+\mu^2\nu^2\big]x^2+(\mu+\nu)\big[1-(\mu+\nu)^2\nonumber\\
&&{}-\mu^2\nu^2(3-2\mu\nu)\big]x^3+\mu\nu(1-\mu\nu)\big[1-(\mu+\nu)^2-\mu^2\nu^2\big]x^4\Big\}\Bigg\}\,.
\ea
\reseteqn
To obtain exactly the form used in \cite{Pomeransky:2006bd}, we have to define the new parameters $\tilde\lambda$, $\tilde\nu$ and $\tilde k$ by
\be
\tilde{\lambda}=\mu+\nu\,,\qquad\tilde{\nu}=\mu\nu\,,\qquad\tilde{k}=\varkappa\,,
\ee
swap the coordinates $\psi$ and $\phi$, and change the signature of the space-time to a mostly minus one.

\newsubsection{Emparan--Reall black ring}

The Emparan--Reall black ring, in the simplest form used in \cite{Emparan:2004wy,Emparan:2008eg}, is obtained by setting $\nu=0$ in (\ref{general_BR}):
\ba
\dif s^2&=&-\frac{\tilde F(y)}{\tilde F(x)}\bigg[\dif t-\varkappa\,\sqrt{\frac{2\lambda(\lambda-\mu)(1+\lambda)}{1-\lambda}}\,\frac{1+y}{\tilde F(y)}\,\dif\psi\bigg]^2+\frac{2\varkappa^2\tilde F(x)}{(x-y)^2}\nonumber\\
&&\quad{}\times\bigg\{-\frac{\tilde G(y)}{\tilde F(y)}\,\dif\psi^2+\frac{\tilde G(x)}{\tilde F(x)}\,\dif\phi^2+\frac{(1-\mu)^2}{1-\lambda}\left[\frac{\dif x^2}{\tilde G(x)}-\frac{\dif y^2}{\tilde G(y)}\right]\bigg\}\,,
\ea
where
\be
\tilde F(x)=1+\lambda x\,,\qquad \tilde G(x)=(1-x^2)(1+\mu x)\,.
\ee
To obtain exactly the form used in \cite{Emparan:2008eg}, we have to define the new parameters $\tilde\lambda$, $\tilde\nu$ and $\tilde R$ by
\be
\tilde\lambda=\lambda\,,\qquad
\tilde\nu=\mu\,,\qquad
\tilde R^2=\frac{2\varkappa^2(1-\mu)^2}{1-\lambda}\,,
\ee
and the new coordinates
\be
(\tilde\psi,\tilde\phi)=\frac{\sqrt{1-\lambda}}{1-\mu}\,(\psi,\phi)\,.
\ee

\newsubsection{Figueras black ring}

The Figueras black ring \cite{Figueras:2005zp} is obtained by setting $\lambda=\mu$ in (\ref{general_BR}):
\ba
\dif s^2&=&-\frac{\tilde H(y,x)}{\tilde H(x,y)}\,\left[\dif t-\varkappa\,\sqrt{\frac{2\mu\nu}{1-\mu\nu}}\,\frac{(\mu+\nu)(1-x^2)y}{\tilde H(y,x)}\,\dif\phi\right]^2+\frac{2\varkappa^2\tilde H(x,y)}{(1-\mu\nu)(x-y)^2}\nonumber\\
&&\quad{}\times\bigg\{-\frac{(1-y^2)\tilde F(x)}{\tilde H(x,y)}\,\dif\psi^2+\frac{(1-x^2)\tilde F(y)}{\tilde H(y,x)}\,\dif\phi^2\nonumber\\
&&\qquad\qquad{}+(1-\mu)(1-\nu)\left[\frac{\dif x^2}{(1-x^2)\tilde F(x)}-\frac{\dif y^2}{(1-y^2)\tilde F(y)}\right]\bigg\}\,,
\ea
where
\be
\tilde H(x,y)=1+(\mu+\nu)x+\mu\nu x^2y^2,\qquad
\tilde F(x)=(1+\mu x)(1+\nu x)\,.
\ee
To obtain exactly the form used in \cite{Figueras:2005zp}, we have to define the new parameters $\tilde\lambda$, $\tilde a$ and $\tilde R$ by
\be
\tilde\lambda=\mu+\nu\,,\qquad
\frac{\tilde a^2}{\tilde R^2}=\mu\nu\,,\qquad
\tilde R^2=\frac{2\varkappa^2(1-\mu)(1-\nu)}{1-\mu\nu}\,,
\ee
and the new coordinates
\be
(\tilde\psi,\tilde\phi)=\frac{1}{\sqrt{(1-\mu)(1-\nu)}}\,(\psi,\phi)\,.
\ee

\newsubsection{Myers--Perry black hole}

The Myers--Perry black hole \cite{Myers:1986un} is obtained by setting $\lambda=1-c(1-\mu)$ for some parameter $0< c\leq1$, performing the coordinate transformation
\ba
x&=&-1+{\frac {8{\varkappa}^{2}  \cos^2\theta \left( 1-\mu \right) }{2{r}^{2}+a^{2}+b^{2}-m-4{\varkappa}^{2} \cos 2\theta}}\,,\cr
y&=&-1-{\frac {8{\varkappa}^{2} \sin^2\theta \left( 1-\mu \right) }{2{r}^{2}+a^{2}+b^{2}-m-4{\varkappa}^{2} \cos 2\theta}}\,,
\ea
where
\ba
m&=&{\frac {4{\varkappa}^{2} \left( 1+\nu \right) }{ c\left(1- \nu \right)}}\,,\cr
a&=&-{\frac { 2\varkappa\sqrt{(1-c^2)(1-\nu)(1+\nu+c-c\nu)} }{\sqrt {c}\left( 1-\nu+c+c\nu \right) }}\,,\cr
b&=&-{\frac {4\varkappa\sqrt {c\nu(1+\nu+c-c\nu)}}{\sqrt {1-\nu}\left( 1-\nu+c+c\nu \right)  }}\,,
\ea
and then taking the limit $\mu\rightarrow1$. If we do this, (\ref{general_BR}) becomes
\ba
\dif s^2&=&-\dif t^2+\frac{m}{\Sigma}(\dif t-a\sin^2\theta\,\dif\psi-b\cos^2\theta\,\dif\phi)^2\cr
&&+(r^2+a^2)\sin^2\theta\,\dif\psi^2+(r^2+b^2)\cos^2\theta\,\dif\phi^2+\Sigma\left(\frac{\dif r^2}{\Delta}+\dif\theta^2\right),
\ea
where
\be
\Delta=r^2\bigg(1+\frac{a^2}{r^2}\bigg)\bigg(1+\frac{b^2}{r^2}\bigg)-m\,,\qquad\Sigma=r^2+a^2\cos^2\theta+b^2\sin^2\theta\,.
\ee
Here, $m$ is the mass parameter of the Myers--Perry black hole, while $a$ and $b$ are the angular-momentum parameters along the $\psi$ and $\phi$ directions, respectively.

\newsubsection{Boosted Kerr black string}

The boosted Kerr black string is obtained by setting
\be
\nu=\frac{m-\sqrt{m^2-a^2}}{\sqrt{2}\varkappa}\,,\qquad
\mu=\frac{m+\sqrt{m^2-a^2}}{\sqrt{2}\varkappa}\,,\qquad
\lambda=\frac{m+\sqrt{m^2-a^2}}{\sqrt{2}\varkappa}\,\cosh\sigma\,,
\ee
changing coordinates $x=\cos\theta$, $y=-\sqrt{2}\varkappa/r$, $\psi=-z/(\sqrt{2}\varkappa)$,
and then sending $\varkappa\rightarrow\infty$. If we do this, (\ref{general_BR}) becomes
\begin{eqnarray}
\dif s^2&=&-\left( 1-\frac{2m r \cosh^2\sigma }{\Sigma}\right) \dif t^2+
\frac{2m r \sinh2\sigma }{\Sigma} \,\dif t\, \dif z + \left( 1+\frac{2m r
\sinh^2\sigma }{\Sigma}\right)\dif z^2\nonumber \\ 
 & & +\frac{(r^2+a^2)^2-\Delta a^2
\sin^2\theta}{\Sigma}\,\sin^2\theta\, \dif \phi^2
  -\frac{4m r \cosh\sigma }{\Sigma}\,a\sin^2\theta\, \dif t\, \dif \phi\nonumber \\
&&  -\frac{4m r \sinh\sigma }{\Sigma}\,
  a\sin^2\theta\, \dif z\, \dif \phi
+\Sigma\left(\frac{\dif r^2}{\Delta}+\dif\theta^2\right),
\end{eqnarray}
where
\be
 \Delta = r^2+ a^2 -2mr \,, \qquad
 \Sigma = r^2+ a^2\cos^2\theta \,.
\ee
This is exactly the metric obtained by starting with the four-dimensional Kerr solution, adding a flat direction $z$ to it, and then applying a boost $\dif t\rightarrow\cosh\sigma\,\dif t+\sinh\sigma\,\dif z$, $\dif z\rightarrow\sinh\sigma\,\dif t+\cosh\sigma\,\dif z$. Some properties of this boosted Kerr black string solution were studied in \cite{Dias:2006zv}.

\bigskip\bigskip

{\renewcommand{\Large}{\normalsize}
}

\end{document}